\documentstyle[amsfonts,12pt,thmsa,sw20aip]{article}

\input tcilatex
\begin{document}

\author{Bobby Eka Gunara\thanks{%
email : bobby@fi.itb.ac.id} \and Theoretical High Energy Physics Group, \and %
Theoretical Physics Laboratory, \and Department of Physics, Bandung
Institute of Technology, \and Jl. Ganesha 10 Bandung, 40132, INDONESIA.}
\title{Semi-Hopf Algebra and Supersymmetry}
\date{THEP-PHYS-ITB-6-99\\
August 1999}
\maketitle

\begin{abstract}
We define a semi-Hopf algebra which is more general than a Hopf algebra.
Then we construct the supersymmetry algebra via the adjoint action on this
semi-Hopf algebra. As a result we have a supersymmetry theory with quantum
gauge group, i.e., quantised enveloping algebra of a simple Lie algebra. For
the example, we construct the Lagrangian ${\cal N}=$1 and ${\cal N}=$2
supersymmetry.
\end{abstract}

\section{Introduction}

Supersymmetry was introduced in the seventy decades by
Golfand-Likht-\thinspace \thinspace man, Volkov-Akulov, Wess-Zumino, and
Salam-Strathdee[1]. This supersymmetry has been the subject of intense
research in particle physics which promises the grand unification theory,
also in field and string theory. The recent development in supersymmetric
field theory is to find a generalized supersymmetry theory via the notion of
duality. It was begin by Seiberg and Witten[3] who considered the ${\cal N}=$%
2 supersymmetric $SU($2$)$ Yang-Mills theory.

On the other hand, Drinfel'd and Jimbo[4] generalized the Lie group (called
quantum group) via noncommutative and non-co-commutative Hopf algebra. This
quantum group has been applied in gauge theory[8,9,10]. This implies that
the field theory become no longer commutative. So it is uneasy to define the
variation of the action and its quantum theory. A different approach
introduced in [14] has successfuly handled this noncommutative problem in
defining the variation of the action.

In this paper, we start with the definition of a semi-Hopf algebra which is
more general than a Hopf algebra, then we construct the supersymmetry
algebra via this semi-Hopf algebra. As a result we have a supersymmetry
theory with quantum gauge group $U_q\left( {\frak g}\right) $, i.e.,
quantised enveloping algebra of a simple Lie algebra ${\frak g}$ which is
defined by Lyubashenko and Sudbery[7]. The field (or superfield) become
noncommutative that we handle by defining a noncommutative invariant scalar
product between two fields (or superfield). So we can construct the
Lagrangian ${\cal N}=$1 and ${\cal N}=$2 supersymmetry by the scalar product
between two superfield and do not worry about the noncommutative factor. It
is also straightforward to define the variation of the action and then get
the usual equation of motion. We use the notation for the supersymmetry as
was given by Wess-Bagger[2].

This paper is organized as follows. In section II we briefly review the
definition of a Hopf algebra given by Tjin[6] and then we define a semi-Hopf
algebra which is more general structure than a Hopf algebra. Then we
construct the supersymmetry algebra via this semi-Hopf algebra and define
the general Jacobi identity in section III. The definition a noncommutative
invariant scalar product and the construction of ${\cal N}=$1 and ${\cal N}=$%
2 supersymmetry Lagrangian via superspace formalism, and the ${\cal N}=$2
prepotential are presented in section IV. Section V is devoted for
conclusion and outlook.

\section{Semi-Hopf Algebra}

Before defining the notion of semi-Hopf algebra, we start with the
definition and elementary properties of Hopf algebra[5]. This Hopf algebra
contains all the information on the algebraic structure of the Lie group
while a semi-Hopf algebra contains all the information on the algebraic
structure of the super-Lie group (as we will see later).

In this paper we use the notation and the definition of a Hopf algebra given
by Tjin[6] and briefly introduce Hopf algebraic structure. Then we define a
semi-Hopf algebra which is more general than a Hopf algebra.

\begin{definition}
Let $A$ is a linear space over the ground field ${\cal C}$. A bi-algebra is
the set $\left( A,m,\Delta ,\eta ,\varepsilon \right) $ with the following
properties :

1) $\left( A,m,\eta \right) $ is unital assosiative algebra, where

\begin{equation}
m:A\otimes A\rightarrow A  \tag{2.1}
\end{equation}

\begin{equation}
\eta :{\cal C}\rightarrow A  \tag{2.2}
\end{equation}

2) A map $\Delta :A\rightarrow A\otimes A$ satisfies

i) $\Delta $ is linear

ii) $\Delta $ is an algebra homomorphism

iii) $\left( \Delta \otimes id\right) \Delta $ $=$ $\left( id\otimes \Delta
\right) \Delta $\qquad (co-associativity)

3) A map $\varepsilon :A\rightarrow {\cal C}$ such that

\begin{equation}
\left( id\otimes \varepsilon \right) \Delta =\left( \varepsilon \otimes
id\right) \Delta =id  \tag{2.3}
\end{equation}
\end{definition}

\begin{definition}
A Hopf algebra is a bi-algebra $\left( A,m,\Delta ,\eta ,\varepsilon \right) 
$ together with a map

\begin{equation}
S:A\rightarrow A  \tag{2.4}
\end{equation}
with the following properties

\begin{equation}
m\left( S\otimes id\right) \Delta =m\left( id\otimes S\right) \Delta =\eta
\circ \varepsilon   \tag{2.5}
\end{equation}
where $S$ is an antipode.
\end{definition}

Some example of a Hopf algebra can be seen in[4-9]. We interested in
constructing an algebra which contain the algebraic structure of the
super-Lie group. So we define an algebra (which we called semi-Hopf algebra)
as follows :

\begin{definition}
A semi-Hopf algebra is a bi-algebra $\left( B,m,\Delta ,\eta ,\varepsilon
\right) $ with the following properties :

1) There exist $q\in B$ such that

\begin{equation}
m\left( id\otimes id\right) \Delta (q)=\eta \circ \varepsilon (q)  \tag{2.6}
\end{equation}

2) There exist $b\in B$ and $b\neq q$ such that

\begin{equation}
m\left( S\otimes id\right) \Delta (b)=m\left( id\otimes S\right) \Delta
(b)=\eta \circ \varepsilon (b)  \tag{2.7}
\end{equation}
where $S$ is an antipode
\end{definition}

We can choose $\eta \circ \varepsilon (a)=\varepsilon (a)\,I_B$ for $a\in B$
with $I_B$ is an identity element of $B$ [4]. Both of the above condition
does not compatible to each other. It means that if $q\in B$ satisfies the
condition (2.6), then $q$ does not satisfy the condition (2.7) and vice
versa.

The definition 3 \thinspace implies the following proposition :

\begin{proposition}
A Hopf algebra is subalgebra of a semi-Hopf algebra.
\end{proposition}

In the next section, we will construct a supersymmetry algebra via a
semi-Hopf algebra where the odd elements, as we will define later, satisfies
the condition (2.6)

\section{Semi-Hopf Algebra and Supersymmetry Algebra}

In this section we start to construct supersymmetry algebra via the
semi-Hopf algebra as we introduced above.

Let $B$ be a bi-algebra generated by $P_\mu $, $J_{\mu \nu }$, $Q_\alpha ^I$%
, $\stackrel{\_}{Q}_{\dot{\alpha}I}$,$E_r$, $F_r$ , $q^{\pm H_r}$, $I$, $%
{\frak J}$ with coproduct $\Delta $, antipode $S$, and counit $\varepsilon $
defined by

\begin{eqnarray}
\Delta \left( P_\mu \right) &=&P_\mu \otimes I+I\otimes P_\mu \text{ , }%
\Delta \left( J_{\mu \nu }\right) =J_{\mu \nu }\otimes I+I\otimes J_{\mu \nu
}\text{ ,}  \tag{3.1} \\
\Delta \left( Q_\alpha ^I\right) &=&Q_\alpha ^I\otimes I+{\frak J}\otimes
Q_\alpha ^I\text{ , }\Delta \left( \stackrel{\_}{Q}_{\dot{\alpha}I}\right) =%
\stackrel{\_}{Q}_{\dot{\alpha}I}\otimes I+{\frak J}\otimes \stackrel{\_}{Q}_{%
\dot{\alpha}I}\text{ ,}  \nonumber \\
\Delta \left( E_r\right) &=&E_r\otimes q^{-H_r}+q^{H_r}\otimes E_r\text{ , }%
\Delta \left( F_r\right) =F_r\otimes q^{-H_r}+q^{H_r}\otimes F_r\text{ ,} 
\nonumber \\
\Delta \left( q^{\pm H_r}\right) &=&q^{\pm H_r}\otimes q^{\pm H_r}\text{, }%
\Delta \left( {\frak J}\right) ={\frak J}\otimes {\frak J}\text{ , }\Delta
\left( I\right) =I\otimes I\text{ ,}  \nonumber
\end{eqnarray}

\begin{eqnarray}
S\left( P_\mu \right) &=&-P_\mu \text{ , }S\left( J_{\mu \nu }\right)
=-J_{\mu \nu }\text{ , }S\left( Q_\alpha ^I\right) =-Q_\alpha ^I\text{ ,} 
\tag{3.2} \\
S\left( \stackrel{\_}{Q}_{\dot{\alpha}I}\right) &=&-\stackrel{\_}{Q}_{\dot{%
\alpha}I}\text{, }S\left( E_r\right) =-q_r^{-1}E_r\text{ , }S\left(
F_r\right) =-q_r\ F_r\text{ ,}  \nonumber \\
S\left( q^{\pm H_r}\right) &=&q^{\mp H_r},S\left( {\frak J}\right) ={\frak J}%
\text{ , }S\left( I\right) =I\text{ ,}  \nonumber
\end{eqnarray}

\begin{eqnarray}
\varepsilon \left( P_\mu \right) &=&\varepsilon \left( J_{\mu \nu }\right)
=\varepsilon \left( Q_\alpha ^I\right) =\text{ 0 ,}  \tag{3.3} \\
\varepsilon \left( \stackrel{\_}{Q}_{\dot{\alpha}I}\right) &=&\varepsilon
\left( E_r\right) =\varepsilon \left( F_r\right) =\text{ 0 ,}  \nonumber \\
\varepsilon \left( q^{\pm H_r}\right) &=&\varepsilon \left( {\frak J}\right)
=\varepsilon \left( I\right) =\text{ 1 ,}  \nonumber
\end{eqnarray}
where $q$ is a fixed elements of ground field ${\cal C}$, $%
q_r=q^{\left\langle H_r\text{ , }H_r\right\rangle }$ and $\left\langle \text{%
,}\right\rangle $ denotes the Killing form in Cartan subalgebra. The
operator ${\frak J}$ acts on the generator of B as

\begin{eqnarray}
{\frak J}Q_\alpha ^I &=&-Q_\alpha ^I\text{ , }{\frak J}\stackrel{\_}{Q}_{%
\dot{\alpha}I}=-\stackrel{\_}{Q}_{\dot{\alpha}I}\text{,}  \tag{3.4} \\
{\frak J}P_\mu &=&P_\mu \text{ , }{\frak J}J_{\mu \nu }=J_{\mu \nu }\text{ , 
}{\frak J}E_r=E_r\text{ ,}  \nonumber \\
{\frak J}F_r &=&F_r\text{ , }{\frak J}q^{\pm H_r}=q^{\pm H_r}\text{ .} 
\nonumber
\end{eqnarray}

It can be seen from the above equation that the basis element of $B$ consist
of the odd elements, i.e., $Q_\alpha ^I$ and $\stackrel{\_}{Q}_{\dot{\alpha}%
I}\,$because they have $-$1 parity and the rest are even, i.e., $P_\mu $, $%
J_{\mu \nu }$, $E_r$, $F_r$ , $q^{\pm H_r}$ because they have $+$1 parity.

The odd elements, i.e., $Q_\alpha ^I$ and $\stackrel{\_}{Q}_{\dot{\alpha}I}$%
, turn out to satisfy the condition (2.6) while the even elements, i.e., $%
P_\mu $, $J_{\mu \nu }$, $E_r$, $F_r$ , $q^{\pm H_r}$, satisfy the condition
(2.7). So it is sufficient for $B$ \thinspace to be a semi-Hopf algebra.

We define the adjoint action of $B$ on itself, given by $x\rightarrow $ ad$%
\,x\in $ End$_{{\frak C}}\,B$ where

\begin{equation}
\text{ ad\thinspace }x\left( y\right) \equiv \sum x_{\left( \text{1}\right)
\,}\,y\,\,S\left( x_{\left( \text{2}\right) }\right) =\left[ x,y\right] 
\text{ ,}  \tag{3.5}
\end{equation}
if $\,\Delta \left( x\right) =\sum x_{\left( \text{1}\right) \,}\otimes
x_{\left( \text{2}\right) \,}$. For odd-odd element the adjoint action become

\begin{eqnarray}
\left[ Q_\alpha ^I,\stackrel{\_}{Q}_{\dot{\beta}J}\right] &=&Q_\alpha ^I%
\stackrel{\_}{Q}_{\dot{\beta}J}+\stackrel{\_}{\text{ \thinspace }Q}_{\dot{%
\beta}J}Q_\alpha ^I\text{ ,}  \tag{3.6} \\
\left[ Q_\alpha ^I,Q_\beta ^J\right] &=&Q_\alpha ^I\,Q_\beta ^J+Q_\beta
^J\,Q_\alpha ^I\,\text{ ,}  \nonumber \\
\left[ \stackrel{\_}{Q}_{\dot{\alpha}I},\stackrel{\_}{Q}_{\dot{\beta}%
J}\right] &=&\stackrel{\_}{Q}_{\dot{\alpha}I}\stackrel{\_}{Q}_{\dot{\beta}J}+%
\stackrel{\_}{\,\,Q}_{\dot{\beta}J}\,\stackrel{\_}{Q}_{\dot{\alpha}I} 
\nonumber
\end{eqnarray}
and for odd-even element (the even element only $P_\mu $, $J_{\mu \nu }$ )
the adjoint action become

\begin{eqnarray}
\left[ Q_\alpha ^I,P_\mu \right] &=&Q_\alpha ^I\,P_\mu -P_\mu \,Q_\alpha ^I\,%
\text{ ,}  \tag{3.7} \\
\left[ Q_\alpha ^I,J_{\mu \nu }\right] &=&Q_\alpha ^I\,J_{\mu \nu }-J_{\mu
\nu }\,Q_\alpha ^I\,\text{ ,}  \nonumber
\end{eqnarray}
and for even-even element (only $P_\mu $, $J_{\mu \nu }$) similar to
equation (3.7). Thus, we can identify $Q_\alpha ^I$ and $\stackrel{\_}{Q}_{%
\dot{\alpha}I}$ as supersymmertry generator, $P_\mu $ and $J_{\mu \nu }$ as
the energy-momentum operators, and the Lorentz rotation generators
(antisymmetric tensor) respectively with $\mu $,$\nu =$0,1,2,3 (spacetime
index with the metric $\eta _{\mu \nu }=$diag$(-$1,1,1,1$)$), $I$ is the
number of supersymmetry generators, and the fermionic index $\alpha (\dot{%
\alpha})=$1$($\.{1}$)$, 2$($\.{2}$)$. The rest even elements, i.e., $E_r$, $%
F_r$ , $q^{\pm H_r}$ are generator of the simply-connected quantised
enveloping algebra $U_q\left( {\frak g}\right) $ and ${\frak g}$ is a simple
Lie algebra[7], where $r$ is the number of fundamental roots of ${\frak g}$
. A quantum Lie algebra ${\frak g}_q$ is a subspace of $U_q\left( {\frak g}%
\right) $ satisfying certain properties[7,8]. So we can construct the
elements of ${\frak g}_q$ (denoted $T^a$ with $a$ is the numbers of basis
element) from $E_r$, $F_r$ , $q^{\pm H_r}$. In this case the coproduct of
the elements of ${\frak g}_q$ are the form [7,8]

\begin{equation}
\Delta \left( T^a\right) =T^a\otimes C+u_b^a\otimes T^b\text{ ,}  \tag{3.8}
\end{equation}
and

\begin{equation}
\text{ad\thinspace }u_b^a\left( T^c\right) =\sigma _{db}^{ac}\,\,T^d\text{ ,}
\tag{3.9}
\end{equation}
where $C$ is a central element of $U_q\left( {\frak g}\right) $ and $\sigma $
is the quantum flip operator \thinspace \thinspace \qquad ( a deformation of
the classical flip operator : $\sigma _{db}^{ac}=\delta _d^c\,\delta _b^a$
when $q=$1).

As we see from equation (3.6) and (3.7), the adjoint action of $B$ have a $%
{\sl Z}_2$ graded structure to preserve the closure property. Thus we can
define the adjoint action on the basis elements of $B$ as

\begin{eqnarray}
\left[ P_\mu ,P_\nu \right] &=&\left[ P_\mu ,T^a\right] =\left[ J_{\mu \nu
},T^a\right] =\text{ 0 ,}  \tag{3.10} \\
\left[ Q_\alpha ^I,\stackrel{\_}{Q}_{\dot{\beta}J}\right] &=&\text{%
2\thinspace }\sigma _{\alpha \dot{\beta}}^\mu \,P_\mu \,\,\delta _J^I\text{ ,%
}  \nonumber \\
\left[ Q_\alpha ^I,Q_\beta ^J\right] &=&\varepsilon _{\alpha \beta }\,Z^{IJ}%
\text{ ,}  \nonumber \\
\left[ \stackrel{\_}{Q}_{\dot{\alpha}I},\stackrel{\_}{Q}_{\dot{\beta}%
J}\right] &=&\varepsilon _{\dot{\alpha}\dot{\beta}}\,Z_{IJ}^{*}\text{ ,} 
\nonumber
\end{eqnarray}

\begin{eqnarray}
\left[ Q_\alpha ^I,Z^{IJ}\right] &=&\left[ \stackrel{\_}{Q}_{\dot{\alpha}%
I},Z^{IJ}\right] =\left[ P_\mu ,Z^{IJ}\right] =\text{ 0 ,}  \tag{3.11} \\
\left[ J_{\mu \nu },Z^{IJ}\right] &=&\left[ T^a,Z^{IJ}\right] =\text{ 0 ,} 
\nonumber \\
\left[ Q_\alpha ^I,P_\mu \right] &=&c\,(\gamma _\mu )_\alpha ^\beta
\,\,\,Q_\beta ^I\text{,}  \nonumber \\
\left[ Q_\alpha ^I,J_{\mu \nu }\right] &=&\,(b_{\mu \nu })_\alpha ^\beta
\,\,\,Q_\beta ^I\text{ ,}  \nonumber
\end{eqnarray}

\begin{eqnarray}
\left[ Q_\alpha ^I,T^a\right] &=&l\left( s^a\right) _J^I\,Q_\alpha ^J\text{ ,%
}  \tag{3.12} \\
\left[ T^a,Q_\alpha ^I\right] &=&r\left( s^a\right) _J^I\,Q_\alpha ^J\text{ ,%
}  \nonumber \\
\left[ Z^{IJ},Z^{IJ}\right] &=&\text{ 0 ,}  \nonumber
\end{eqnarray}

\begin{eqnarray}
\left[ T^a,T^b\right] &=&f_c^{ab}\,T^c\text{ ,}  \tag{3.13} \\
\left[ P_\mu ,J_{\rho \nu }\right] &=&\eta _{\mu \rho }\,P_\nu -\eta _{\mu
\nu }\,P_\rho \text{ ,}  \nonumber \\
\left[ J_{\mu \nu },J_{\rho \sigma }\right] &=&\eta _{\mu \sigma }\,J_{\nu
\rho }+\eta _{\nu \rho }\,J_{\mu \sigma }-\eta _{\mu \rho }\,J_{\nu \sigma
}-\eta _{\nu \sigma }\,J_{\mu \rho }\text{ ,}  \nonumber
\end{eqnarray}
where $c$ is a constant, $f_c^{ab}$ is a structure constant, $Z^{IJ}$ are
the central charge with its conjugate $Z_{IJ}^{*}$ , $\sigma ^\mu $ are the
Pauli matrices, $\varepsilon _{\alpha \beta }$ and $\varepsilon _{\stackrel{%
\cdot }{\alpha }\stackrel{\cdot }{\beta }}$ are the Levi-Civita symbol, $l$
and $r$ denote left and right , and $\gamma _\mu $ and $b_{\mu \nu }$ are
matrix vector and matrix antisymmetric tensor, respectively. The central
charges $Z^{IJ}$ are even elements of $B$. The set of the adjoint action in
equation (3.10) through (3.13) is called general supersymmetry algebra. The
second and thirth equation in equation (3.10) are set to be zero in order to
preserve the Coleman-Mandula theorem[12].

Let $x_{\stackunder{-}{\text{1}}}=P_\mu $ , $x_{\stackunder{-}{\text{2}}%
}=J_{\mu \nu }$ , $x_{\stackunder{-}{\text{3}}}=Q_\alpha ^I$ , $x_{%
\stackunder{-}{\text{4}}}=\stackrel{\_}{Q}_{\dot{\beta}J}$, $x_a=T^a$ with $%
a=$ 1,..., $N$ and $N$ is the numbers of generator of the gauge group. The
general supersymmetry algebra satisfies :

1. The left Jacobi identity

\begin{equation}
\left[ \left[ x_{\stackunder{-}{i}},x_{\stackunder{-}{j}}\right] ,x_{%
\stackunder{-}{k}}\right] =\Gamma _{_{\stackunder{-}{i\,}\stackunder{-}{j}%
}}^{_{\stackunder{-}{l\,}\stackunder{-}{m}}}\left[ x_{\stackunder{-}{l}%
}\left[ x_{\stackunder{-}{m}},x_{\stackunder{-}{k}}\right] \right] \text{ .}
\tag{3.14}
\end{equation}

2. The right Jacobi identity

\begin{equation}
\left[ x_{\stackunder{-}{i}}\left[ x_{\stackunder{-}{j}},x_{\stackunder{-}{k}%
}\right] \right] =\Gamma _{_{\stackunder{-}{j\,}\stackunder{-}{k}}}^{_{%
\stackunder{-}{l\,}\stackunder{-}{m}}}\left[ \left[ x_{\stackunder{-}{i}},x_{%
\stackunder{-}{l}}\right] ,x_{\stackunder{-}{m}}\right] \text{ .}  \tag{3.15}
\end{equation}
$\Gamma _{_{\stackunder{-}{j\,}\stackunder{-}{k}}}^{_{\stackunder{-}{l\,}%
\stackunder{-}{m}}}$ is the antisymmetriser with $\stackunder{-}{i}$%
,\thinspace $\stackunder{-}{j}$ , $\stackunder{-}{k}$ , $\stackunder{-}{l}$
, $\stackunder{-}{m}$ $=$ $\stackunder{-}{\text{1}}$ , $\stackunder{-}{\text{%
2}}$ , $\stackunder{-}{\text{3}}$ , $\stackunder{-}{\text{4}}$ , $a$,
\thinspace which will reduce to the antisymmetriser defined by
Lyubashenko-Sudbery[7] \thinspace if $\stackunder{-}{l}=a,\,\stackunder{-}{m}%
=b,\stackunder{-}{\,j}=c,\stackunder{-}{k}=d$ (where $a,b,c,d=$ 1,...,$N$) ,
i.e., $\Gamma _{cd}^{ab}=\gamma _{cd}^{ab}$ .

The antisymmetriser $\Gamma $ is restricted by the general supersymmetry
algebra \thinspace \thinspace (equation (3.10) through (3.13)) for which $c= 
$ 0, $b_{\mu \nu }$ form a representation of Lorentz algebra, and $l\left(
s^a\right) $, $r\left( s^a\right) $ represent of internal symmetry algebra.

\section{$U_q\left( {\frak g}\right) $ Supersymmetry}

In the previous section, we have the general supersymmetry algebra\thinspace
\qquad \thinspace \quad ( equation (3.10) through (3.13)), i.e., the
supersymmetry algebra with quantum gauge group $U_q\left( {\frak g}\right) $%
. This algebra implies that the massless and massive mutiplet are precisely
the same as the supersymmetry with classical gauge group. The difference is
that in this theory we have noncommutative fields (or superfields) in each
multiplet. So we need to define the noncommutative factor between two
different fields (or superfields). Some example for the gauge theory are in
references [8,9,10]. It is a complicated problem to define the variation of
action of the theory. Fortunately, the noncomutative invariant property of
the scalar product between two fields ( or superfields) bring us out of this
problem.

\subsection{Noncommutative factor and Metric}

\begin{definition}
Let $\Phi ^{\bar{a}_1}$ and $\Psi ^{\bar{b}_2}$ are two different kind of
field (or superfield) in $\rho _1\,$and $\rho _2$ representation,
respectively. Then there exist a factor $B_{\bar{c}_2\bar{d}_1}^{\bar{a}_1%
\bar{b}_2}$ such that

\begin{equation}
\Phi ^{\bar{a}_1}\,\Psi ^{\bar{b}_2}=B_{\bar{c}_2\bar{d}_1}^{\bar{a}_1\bar{b}%
_2}\,\Psi ^{\bar{c}_2}\Phi ^{\bar{d}_1}  \tag{4.1}
\end{equation}
$B_{\bar{c}_2\bar{d}_1}^{\bar{a}_1\bar{b}_2}$ is called the noncommutative
factor.

$\,$
\end{definition}

The noncommutative factor $B_{\bar{c}_2\bar{d}_1}^{\bar{a}_1\bar{b}_2}$ is
an $Nn$ $\times \,Nn$ matrix with $N$ is the dimension of $\rho _1$
representation and $n$ is the dimension of $\rho _2$ representation.

If $\Phi $ and $\Psi $ have the same representation, i.e. , $\rho _1$ $=\rho
_2=\rho $ then the noncommutative factor become

\begin{equation}
\Phi ^{\bar{a}}\,\Psi ^{\bar{b}}=\lambda _{\rho ,\,\bar{c}\bar{d}}^{\bar{a}%
\bar{b}}\,\Psi ^{\bar{c}}\Phi ^{\bar{d}}\text{ ,}  \tag{4.2}
\end{equation}
where $\lambda _{\rho ,\bar{c}\bar{d}}^{\bar{a}\bar{b}}$ is an $N^2\times
N^2 $ matrix with $N$ is the dimension of the $\rho $ representation. For
special case, if $\rho $ is an adjoint representation then $\lambda _{\text{%
ad},\,cd}^{ab}=\sigma _{cd}^{ab}$, where $\sigma $ is the quantum flip
operator defined in [7,8].

\begin{definition}
Let $\Phi $ and $\Psi $ are two different kind of fields (or superfields) in
the $\rho $ representation. The product of two fields (or superfields)

\[
g_{\bar{a}\bar{b}}\,\,\Phi ^{\bar{a}}\,\Psi ^{\bar{b}}\equiv \Phi ^{\bar{a}%
}\,\Psi _{\bar{a}} 
\]
is called a noncommutative invariant scalar product if it satisfies

\begin{equation}
\Phi ^{\bar{a}}\,\Psi _{\bar{a}}=\Psi ^{\bar{a}}\,\Phi _{\bar{a}}\text{ .} 
\tag{4.3}
\end{equation}
\end{definition}

In the rest of this paper, the noncommutative invariant scalar product will
be simply called the scalar product. The function $g_{\bar{a}\,\bar{b}}$ has
form

\begin{equation}
g_{\bar{a}\bar{b}}=\text{Tr}\left( u\,\,e_{\bar{a}}\,\,e_{\bar{b}}\right) 
\text{ ,}  \tag{4.4}
\end{equation}
where $e_{\bar{a}}$ and $\,e_{\bar{b}}$ are basis of the $\rho $
representation, $u=\sum S\left( {\cal R}_2\right) {\cal R}_1$ being the
quantum trace element of $U_q\left( {\frak g}\right) $ and ${\cal R}=\sum 
{\cal R}_1\otimes {\cal R}_2$ its universal R-matrix. We have that the
function $g_{\bar{a}\,\bar{b}}$ contracting the upper index to the lower
index, that is

\begin{equation}
g_{\bar{a}\bar{b}}\,\Phi ^{\bar{b}}=\Phi _{\bar{a}}\text{ ,}  \tag{4.5}
\end{equation}
and vice versa. The function $g_{\bar{a}\,\bar{b}}$ also satisfies

\begin{equation}
g^{\bar{a}\bar{b}}g_{\,\bar{b}\bar{c}}=\delta _{\bar{c}}^{\bar{a}}\text{ .} 
\tag{4.6}
\end{equation}

\begin{proposition}
The function $g_{\bar{a}\,\bar{b}}$ satisfies $g_{\bar{a}\,\bar{b}}\,\lambda
_{\rho ,\,\bar{c}\bar{d}}^{\bar{a}\bar{b}}=g_{\bar{c}\bar{d}}$ .
\end{proposition}

For the adjoint representation, we have

\[
g_{ab}=\text{Tr\thinspace }\left( u\,\text{\thinspace ad}T^a\,\text{ad}%
T^a\right) \text{ ,} 
\]
and for the $SU_q\left( \text{2}\right) $ see Sudbery[7,8].

Note that it is straightforward to define the variation of the
Lagrangi-\thinspace \thinspace \quad an : Let ${\cal L}\left( \Psi ,\Psi
_{,\mu }\right) $ be a Lagrangian, then we define the variation of ${\cal L}$
as

\begin{equation}
\delta {\cal L=}\delta \Psi ^{\bar{a}}\,\frac{\partial {\cal L}}{\partial
\Psi ^{\bar{a}}}+\delta \Psi _{,\mu }^{\bar{a}}\frac{\partial {\cal L}}{%
\partial \Psi _{,\mu }^{\bar{a}}}\text{ ,}  \tag{4.7}
\end{equation}
where $\Psi $ is a field and $\Psi _{,\mu }=\partial \Psi /\partial x^\mu $.
We cancel the surface term and then get the usual equation of motion.

\subsection{${\cal N}=$1 Scalar and Vector Multiplet}

\subsubsection{${\cal N}=$1 Scalar Multiplet}

The ${\cal N}=$1 scalar multiplet is represented by a chiral superfield $%
\Phi =\Phi ^{\bar{a}}e_{\bar{a}}$ and can be expanded as

\begin{equation}
\Phi \left( y,\theta \right) =\phi \left( y\right) +\sqrt{2}\,\,\,\theta
\psi \left( y\right) +\theta ^2\,F\left( y\right) \text{ ,}  \tag{4.8}
\end{equation}
where $y^\mu =x^\mu +i\theta \,\sigma ^\mu \,\bar{\theta}$, $x$ are
spacetime coordinates, $\theta \,,\,\bar{\theta}$ are anticommuting
coordinates, and $\sigma ^\mu $ are the Pauli matrices. Here, $\phi $ and $%
\psi $ are the scalar and fermionic components respectively and $F$ is an
auxiliary field required for the off-shell representation.

Similarly, an anti-chiral superfield is represented by $\bar{\Phi}=\Phi
^{\dagger ,\,\bar{a}}e_{\bar{a}}$ and can be expanded as

\begin{equation}
\bar{\Phi}\left( y^{\dagger },\bar{\theta}\right) =\bar{\phi}\left(
y^{\dagger }\right) +\sqrt{2}\,\,\,\bar{\theta}\bar{\psi}\left( y^{\dagger
}\right) +\bar{\theta}^2\,\bar{F}\left( y^{\dagger }\right) \text{ ,} 
\tag{4.9}
\end{equation}
where $y^{\dagger ,\mu }=x^{\dagger ,\mu }-i\theta \,\sigma ^\mu \,\bar{%
\theta}$ .

The Lagrangian for the scalar multiplet is the scalar product of a chiral
and an anti-chiral superfield

\begin{equation}
{\cal L}=\int \Phi ^{\dagger ,\,\bar{a}}\Phi _{\bar{a}}\,\,d^{\text{4}%
}\theta =(\partial _\mu \phi ^{\dagger ,\,\bar{a}})(\partial ^\mu \phi _{%
\bar{a}})-i\bar{\psi}^{\bar{a}}\sigma ^\mu \,\partial _\mu \psi +F^{\dagger
,\,\bar{a}}\,F_{\bar{a}}\text{ ,}  \tag{4.10}
\end{equation}

where $d^{\text{4}}\theta =d^{\text{2}}\theta \,\,d^{\text{2}}\bar{\theta}$ .

\subsubsection{${\cal N}=$1 Vector Multiplet}

This multiplet is represented by a real superfield satisfying $V^{\dagger
}=V $. Using the abelian gauge transformation

\begin{equation}
V\rightarrow V+a\,\Lambda +b\,S\left( \Lambda ^{\dagger }\right) \text{ ,} 
\tag{4.11}
\end{equation}
where $\Lambda (\Lambda ^{\dagger })$ are chiral (anti-chiral) superfield, $%
a,b$ are coefficient, and $S$ is an antipode, we can write $V$ as

\begin{equation}
V=-\theta \,\sigma \,\bar{\theta}\,A_\mu +i\theta ^2\,\bar{\theta}\bar{%
\lambda}-i\bar{\theta}^2\,\theta \lambda +\frac{\text{1}}{\text{2}}\theta
^2\,\bar{\theta}^2\,D\text{ ,}  \tag{4.12}
\end{equation}
the so called Wess-Zumino gauge. The abelian field strength is defined by

\begin{equation}
W_\alpha =-\frac{\text{1}}{\text{4}}\bar{D}^2D_\alpha V\text{ , }\bar{W}_{%
\dot{\alpha}}=-\frac{\text{1}}{\text{4}}D^2\bar{D}_{\dot{\alpha}}V\text{ ,} 
\tag{4.13}
\end{equation}
where $D^2=D^\alpha D_\alpha $ and $\bar{D}^2=\bar{D}_{\dot{\alpha}}\bar{D}^{%
\dot{\alpha}}$, and $D^\alpha $, $\bar{D}_{\dot{\alpha}}$ are the
super-covariant derivatives[2]. $W_\alpha $ is a chiral superfield.

In the non-abelian case, $V$ belongs to the adjoin representation of the
gauge group $U_q\left( {\frak g}\right) $ : $V=V_a$ ad$\,T^a$ . The gauge
transformations are now implemented by

\begin{eqnarray}
e^{-\text{2}V} &\rightarrow &\sum h_{\left( \text{1}\right) }\left( \bar{%
\Lambda}\right) \,e^{-\text{2}V}\,S\left( h_{\left( \text{2}\right) }\left(
\Lambda \right) \right) \text{ ,}  \tag{4.14} \\
e^{\text{2}V} &\rightarrow &\sum h_{\left( \text{1}\right) }\left( \Lambda
\right) \,e^{\text{2}V}\,S\left( h_{\left( \text{2}\right) }\left( \bar{%
\Lambda}\right) \right) \text{ ,}  \nonumber
\end{eqnarray}
if $\Delta \left( h\right) =\sum h_{\left( \text{1}\right) }\otimes
h_{\left( \text{2}\right) }$ , where $\Lambda =\Lambda _a\,$ad$\,T^a$ and $%
\bar{\Lambda}=\Lambda _a^{\dagger }\,$ad$\,T^a$ . The non-abelian gauge
field strength is defined by

\begin{equation}
W_\alpha =\frac{\text{1}}{\text{8}}\bar{D}^2\,e^{\text{2}V}D_\alpha \,e^{-%
\text{2}V}\text{ , }\bar{W}_{\dot{\alpha}}=\frac{\text{1}}{\text{8}}D^2e^{%
\text{2}V}\bar{D}_{\dot{\alpha}}\,e^{-\text{2}V}\text{ ,}  \tag{4.15}
\end{equation}
and as we expect they transform as

\begin{equation}
W_\alpha ^{\prime }\rightarrow \sum h_{\left( \text{1}\right) }\left(
\Lambda \right) \,W_\alpha \,S\left( h_{\left( \text{2}\right) }\left(
\Lambda \right) \right) \text{ .}  \tag{4.16}
\end{equation}
In components, it takes the form

\begin{equation}
W_\alpha =\left( -i\lambda _{a,\alpha \,}+\theta _\alpha \,D_a-\frac i{\text{%
2}}(\sigma ^\mu \,\bar{\sigma}^\nu \theta )_\alpha \,F_{a,\,\mu \nu }+\theta
^2(\sigma ^\mu \,\nabla _\mu \,\bar{\lambda}_a)_\alpha \right) \,\text{%
ad\thinspace }T^a\text{ ,}  \tag{4.17}
\end{equation}
where

\begin{eqnarray*}
F_{\,\mu \nu }^a &=&\partial _\mu \,A_\nu ^a-\partial _\nu \,A_\mu
^a+C\,f_{bc}^a\,\,A_\mu ^b\,\;\text{\/}A_\nu ^c\text{ ,} \\
\nabla _\mu \,\bar{\lambda}^a &=&\partial _\mu \,\bar{\lambda}%
^a+C\,f_{bc}^a\,\,A_\mu ^b\,\bar{\lambda}^c\text{ ,}
\end{eqnarray*}
and $C$ is the central element of $U_q\left( {\frak g}\right) $.

The non-abelian supersymmetric Lagrangian (with the $F\tilde{F}$-term) is
given by

\begin{equation}
{\cal L}=\frac{\text{1}}{\text{8}\pi }\text{ Im}\left( \tau \text{\/}\int d^{%
\text{2}}\theta \text{\/\/\thinspace }W^{a,\,\alpha \,}W_{a,\,\alpha
}\right) \text{ ,}  \tag{4.18}
\end{equation}
where $\tau =\frac \theta {\text{2}\pi }+\frac{\text{4}\pi i}{g^2}$ (see
[13]), $\theta $ is a real parameter, and $g$ is a coupling constant.

\subsubsection{${\cal N}=$1 Vector Multiplet couple with Matter}

Let $\Phi $ be a chiral superfield belong to a $\rho $ representation of the
gauge group $U_q\left( {\frak g}\right) $. The kinetic term $\Phi ^{\dagger
,\,\bar{a}}\Phi _{\bar{a}}$ is invariant under the global gauge
transformation $\Phi ^{\prime }=h(\Lambda )\,\Phi $ and $\bar{\Phi}^{\prime
}=\,\bar{\Phi}\,S\left( h(\Lambda )\right) $ with $\Lambda $ a real
parameter. In the local case, to insure that $\Phi ^{\prime }$ and $\bar{\Phi%
}^{\prime }$ remain a chiral and anti-chiral superfield, $\Lambda $ and $%
\bar{\Lambda}$ have to be a chiral and anti-chiral superfield, respectively.
So the local gauge transformation of a chiral and an anti-chiral superfield
become

\begin{eqnarray}
\Phi ^{\prime } &=&h(\Lambda )\,\Phi \text{ ,}  \tag{4.19} \\
\bar{\Phi}^{\prime } &=&\,\bar{\Phi}\,S\left( h(\bar{\Lambda})\right) \text{
,}  \nonumber
\end{eqnarray}
respectively. The supersymmetric gauge invariant kinetic term is given by

\begin{equation}
\bar{\Phi}\,\,e^{-\text{2}gV}\,\Phi \text{ .}  \tag{4.20}
\end{equation}
Then the Lagrangian ${\cal N}=$1 vector multiplet couple with the kinetic
matter term is

\begin{equation}
{\cal L}=\frac{\text{1}}{\text{8}\pi }\text{ Im}\left( \tau \text{\/}\int d^{%
\text{2}}\theta \text{\/\/\thinspace }W^{a,\,\alpha \,}W_{a,\,\alpha
}\right) +\int d^{\text{4}}\theta \text{ }\bar{\Phi}\,\,e^{-\text{2}%
gV}\,\Phi \text{ .}  \tag{4.21}
\end{equation}
In terms of the superfield components, the above Lagrangian takes the form

\begin{eqnarray}
{\cal L} &=&-\frac{\text{1}}{\text{4}g^2}F_{\mu \nu }^a\,F_a^{\mu \nu
}+\frac \theta {\text{32}\pi ^{\text{2}}}F_{\mu \nu }^a\,\tilde{F}_a^{\mu
\nu }-\frac i{\text{2}g^{\text{2}}}\lambda ^a\,\sigma ^\mu \,\nabla _\mu \,%
\bar{\lambda}_a  \tag{4.22} \\
&&+\frac i{\text{2}g^{\text{2}}}\bar{\lambda}^a\,\sigma ^\mu \,\nabla _\mu
\,\lambda _a+\frac{\text{1}}{\text{2}g^{\text{2}}}D^aD_a+(\nabla _\mu
^{(\rho )}\bar{\phi})(\nabla ^{(\rho ),\,\mu }\phi )  \nonumber \\
&&-i\bar{\psi}\,\sigma ^\mu \,\nabla _\mu ^{(\rho )}\,\psi -\bar{\phi}D\phi
-i\sqrt{2}\,\bar{\phi}\lambda \psi +i\sqrt{2}\,\bar{\psi}\bar{\lambda}\phi +%
\bar{F}F\text{ ,}  \nonumber
\end{eqnarray}
where

\[
\nabla _\mu ^{(\rho )}\bar{\phi}=\partial _\mu \bar{\phi}+ig\,\bar{\phi}%
\,A_\mu \text{ ,}\quad \nabla _\mu ^{(\rho )}\phi =\partial _\mu \phi
+ig\,A_\mu \,\phi \text{ .} 
\]
The above Lagrangian is invariant under the supersymmetry variations, $%
\delta _\epsilon =\epsilon ^\alpha \,Q_\alpha +\bar{\epsilon}_{\dot{\alpha}%
}\,\bar{Q}^{\dot{\alpha}}$ , i.e.,

\begin{eqnarray}
\delta _\epsilon \,A_\mu &=&i\lambda ^a\,\sigma ^\mu \,\bar{\epsilon}%
-i\,\epsilon \,\sigma ^\mu \,\bar{\lambda}^a\text{ ,}  \tag{4.23} \\
\delta _\epsilon \,D^a &=&-\,\nabla _\mu \,\lambda ^a\,\sigma ^\mu \,\bar{%
\epsilon}-\epsilon \,\sigma ^\mu \,\nabla _\mu \,\bar{\lambda}^a\text{ ,} 
\nonumber \\
\delta _\epsilon \,\lambda ^a &=&\sigma ^{\mu \nu \,}\epsilon \,\,F_{\mu \nu
}^a+i\,\epsilon \,D^a\text{ ,}  \nonumber \\
\delta _\epsilon \,\bar{\lambda}^a &=&\bar{\epsilon}\,\bar{\sigma}^{\nu \mu
\,}\,\,F_{\mu \nu }^a-i\,\bar{\epsilon}\,\,D^a\text{ ,}  \nonumber \\
\delta _\epsilon \,\phi &=&\sqrt{2}\,\epsilon \psi \text{ ,}  \nonumber \\
\delta _\epsilon \,\psi &=&\sqrt{2}\,\epsilon \,F+i\sqrt{2}\,\,\sigma ^\mu \,%
\bar{\epsilon}\,\nabla _\mu ^{(\rho )}\,\phi \text{ ,}  \nonumber \\
\delta _\epsilon \bar{\psi} &=&\sqrt{2}\,\bar{\epsilon}\,\bar{F}-i\sqrt{2}%
\,\epsilon \,\sigma ^\mu \,\,\nabla _\mu ^{(\rho )}\,\bar{\phi}\text{ ,} 
\nonumber \\
\delta _\epsilon \,F &=&-i\sqrt{2}\,\nabla _\mu ^{(\rho )}\,\psi \,\,\sigma
^\mu \,\,\bar{\epsilon}-\text{2}i\phi \,\bar{\epsilon}\,\bar{\lambda}\,\text{
.}  \nonumber
\end{eqnarray}
We see from the above equation that the supersymmetry variations of field in
this theory are precisely the same as the classical internal symmetry group
of supersymmetry[2].

\subsection{${\cal N}=$2 Vector Multiplet}

${\cal N}=$2 vector multiplet consist of fields $\phi ,\psi $ and $A_{\mu
\,},\lambda $ in a single multiplet. This means that all fields must be in
the same representation of the gauge group $U_q\left( {\frak g}\right) $ as $%
A_\mu $, i.e., in the adjoint representation. So, the Lagrangian ${\cal N}=$%
2 vector multiplet is the Lagrangian (4.21)( or (4.22)) with the matter
multiplet in the adjoint representation, i.e.,

\begin{equation}
{\cal L}=\text{ Im}\left( \frac \tau {\text{8}\pi }\text{\/}\left[ \int d^{%
\text{2}}\theta \text{\/\/\thinspace }W^{a,\,\alpha \,}W_{a,\,\alpha }+\text{%
2}\int d^{\text{4}}\theta \text{ }\bar{\Phi}\,\,e^{-\text{2}gV}\,\Phi
\right] \right) \text{ ,}  \tag{4.24}
\end{equation}
with scaling $\Phi \rightarrow \Phi /g$ .

The Lagrangian ${\cal N}=$2 vector multiplet can be constructed by using $%
{\cal N}=$2 superspace formalism[11]. The ${\cal N}=$2 chiral superfield is
introduced as follows

\begin{equation}
\Psi =\Phi \left( \tilde{y},\theta \right) +\sqrt{\text{2}}\tilde{\theta}%
^\alpha \,W_\alpha \left( \tilde{y},\theta \right) +\tilde{\theta}%
^2\,G\left( \tilde{y},\theta \right) \text{ ,}  \tag{4.25}
\end{equation}
where $\tilde{y}^\mu =x^\mu +i\,\theta \,\sigma ^\mu \,\bar{\theta}+i\,%
\tilde{\theta}\,\sigma ^\mu \,$\/$\,\!\!$\/$\stackrel{-}{\text{%
\/\negthinspace }\tilde{\theta}}$ and $\tilde{\theta}\,,\,\stackrel{-}{%
\tilde{\theta}}$ are addition anti-commuting coordinates. Then the
Lagrangian (4.24) can be written as

\begin{equation}
{\cal L}=\frac{\text{1}}{\text{4}\pi }\text{ Im}\left( \int d^{\text{2}%
}\theta \text{\/\/\thinspace }d^{\text{2}}\tilde{\theta}\text{\/\/\thinspace
\thinspace }\frac{\text{1}}{\text{2}}\tau \,\Psi ^a\,\Psi _a\text{\/}\right) 
\text{ ,}  \tag{4.26}
\end{equation}
with

\begin{equation}
G\left( \tilde{y},\theta \right) =\int d^{\text{2}}\bar{\theta}\,\bar{\Phi}%
\left( \tilde{y}^{\mu ,\dagger }+i\theta \,\sigma ^\mu \,\bar{\theta},\theta
,\bar{\theta}\right) \text{exp}\left( -\text{2}g\,V\left( \tilde{y}^\mu
-i\theta \,\sigma ^\mu \,\bar{\theta},\theta ,\bar{\theta}\right) \right) 
\text{ .}  \tag{4.27}
\end{equation}
The most general Lagrangian for ${\cal N}=$2 vector multiplet is

\begin{eqnarray}
{\cal L} &=&\frac{\text{1}}{\text{4}\pi }\text{ Im}\left( \int d^{\text{2}%
}\theta \text{\/\/\thinspace }d^{\text{2}}\tilde{\theta}\text{\/\/\thinspace
\thinspace }{\cal F}\left( q,\Psi \right) \,\text{\/}\right) \text{ ,} 
\tag{4.28} \\
&=&\frac{\text{1}}{\text{8}\pi }\text{Im}\left( \text{\/}\left[ \int d^{%
\text{2}}\theta \text{\/\/\thinspace }W^{a,\,\alpha \,}W_{\,\alpha }^b\,%
{\cal F}_{ab}\left( q,\Phi \right) +\text{2}\int d^{\text{4}}\theta \text{ }%
\left( \bar{\Phi}\,\,e^{-\text{2}gV}\right) ^a\,{\cal F}_a\left( q,\Phi
\right) \right] \right) \text{ .}  \nonumber
\end{eqnarray}
Here, ${\cal F}_a\left( q,\Phi \right) =\partial {\cal F}/\partial \Phi ^a$, 
${\cal F}_{ab}\left( q,\Phi \right) =\partial ^2{\cal F}/\partial \Phi
^a\partial \Phi ^b$, and ${\cal F}$ is referred to as the ${\cal N}=$2
prepotential[3]. The second term in the above equation has the K\"{a}hler
Potential Im$\left( \bar{\Phi}\,^a\,{\cal F}_a\left( q,\Phi \right) \right) $
which gives a metric on the space of fields. Seiberg and Witten[3] have
determined exactly this function ${\cal F}$ for the $SU$(2) gauge group.

\section{Conclusion and Outlook}

We have defined a semi-Hopf algebra which is more general than a Hopf
algebra and we construct the supersymmetry algebra via the adjoint action on
this semi-Hopf algebra. As a result we have a supersymmetry theory with
quantum gauge group $U_q\left( {\frak g}\right) $, i.e., quantised
enveloping algebra of a simple Lie algebra ${\frak g}$ . The field (or
superfield) become noncommutative that we handled by defining the
noncommutative invariant scalar product between two fields (or superfield).
We then construct the Lagrangian of the ${\cal N}=$1 and ${\cal N}=$2
supersymmetry by the scalar product between two superfield and do not worry
about the noncommutative factor. We find that the ${\cal N}=$2 prepotential
depends on a fixed element $q$ of the ground field ${\cal C}$ and the chiral
superfield $\Phi $. We left the problems of finding R-matrix for the
quasi-semi-Hopf algebra and the extended Seiberg-Witten theory of quantum
gauge group for further research.

\section{Acknowledgement}

I would like to thank A. Sudbery, F. P. Zen, R. Muhamad, D. P. Hutasoit, and
H. P. Handoyo for discussion and H. J. Wospakrik for useful comments. I also
thank P.Silaban for his encouragement.

\end{document}